\documentclass{llncs}
\usepackage{epstopdf}
\usepackage{multirow}
\usepackage{mathtools}
\usepackage{subfloat}
\usepackage{subfigure}
\usepackage{graphicx}
\usepackage{stfloats}
\usepackage{enumitem}
\usepackage{fancyvrb}
\usepackage{color}

\begin{document}
\clubpenalty=10000 
\widowpenalty = 10000 

\title{Carbon Dating The Web:\\ Estimating the Age of Web Resources}

\author{Hany M. SalahEldeen and Michael L. Nelson}
\institute{Old Dominion University, Department of Computer Science\\
Norfolk VA, 23529, USA\\
{hany,mln}@cs.odu.edu}

\maketitle
\begin{abstract}
In the course of web research it is often necessary to estimate the 
creation datetime for web resources (in the general case, this value can 
only be estimated).  While it is feasible to manually establish likely 
datetime values for small numbers of resources, this becomes infeasible if 
the collection is large.  We present ``carbon date'', a simple web 
application that estimates the creation date for a URI by polling a number 
of sources of evidence and returning a machine-readable structure with their 
respective values.  To establish a likely datetime, we poll bitly for the 
first time someone shortened the URI, topsy for the first time someone 
tweeted the URI, a Memento aggregator for the first time it appeared in a 
public web archive, Google's time of last crawl, and the Last-Modified HTTP 
response header of the resource itself.  We also examine the backlinks of 
the URI as reported by Google and apply the same techniques for the resources 
that link to the URI.  We evaluated our tool on a gold standard data set 
of 1200 URIs in which the creation date was manually verified.  We were 
able to estimate a creation date for 75.90\% of the resources, with 32.78\% 
having the correct value.  Given the different nature of the URIs, the 
union of the various methods produces the best results.  While the Google 
last crawl date and topsy account for nearly 66\% of the closest answers, 
eliminating the web archives or Last-Modified from the results produces 
the largest overall negative impact on the results.  The carbon date application 
is available for download or use via a web API.

\end{abstract}



\section{Introduction}
On numerous occasions during our research in social media, resource sharing, intention analysis, and dissemination patterns, an interesting question emerged: When did a certain resource first appear on the public web? Upon examining a resource, one could find a publishing timestamp indicating when this 
resource was created or first made available to the public. For those select few pages, the timestamp format varies largely along with the time granularity. Some forum posts could deliver solely the month 
and the year of publishing, while in other news sites one can extract the timestamp to the second. Time zones could be problematic too: if not clearly stated on the page, the time zone could be that of the webserver, crawler/archive, or GMT.

Ideally, each resource should be accompanied by a creation date timestamp but this is not true in most cases. A second resort would be to ask the hosting web server to return the last modified HTTP response header. 
Unfortunately, a large number of servers deliberately return more current last modified dates to persuade the search engine crawlers to continuously crawl the hosted pages. This renders the dates obtained 
from the resource or its server highly unreliable.

In our prior work, some of the social media resources we were investigating, ceased to exist. We needed to investigate the time line of this resource from creation, 
to sharing, to deletion. Depending on the hosting server to provide historic information about a missing resource is unachievable in most cases. This places a limitation to services that attempt to parse the 
resource textual representation or even its URI looking for timestamps.

The following step would be to search the public archives for the first existence of the resource. As we show below that using this method solely has significant limitations.

\begin{figure*}[hb]
\centering
\includegraphics[width=\textwidth]{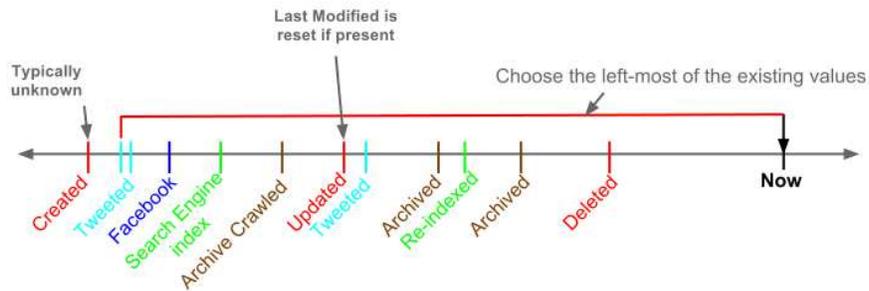}
\caption{The timeline of a shared resource and the proposed process of carbon dating}
\label{fig:timeline}
\end{figure*}
Thus there is a need for a tool that can estimate the creation date of any resource investigated without relying on the infrastructure of the hosting web server or the state of the resource itself. 
Some pages are associated with APIs or tools to extract its metadata, but unfortunately they are non-unified, extremely specific, and what works on one page would not necessarily work on the other.

Due to the speed of web content creation and the ease of publishing, a certain assumption could be established. In some cases, like in blogs, a page could be created and edited before it is published to the public. To facilitate our analysis, 
we will assume that the creation and publishing of a resource coincide. If the creation 
date of the resource is unattainable, then the timestamp of its publishing or release could suffice as a fairly accurate estimate of the creation date of the resource. As fire leaves traces of smoke and ashes, web resources leave traces in references, likes, and backlinks. The 
events associated with creating those shares, links, likes, and interaction with the URI could act as an estimate as well. If we have access to these events, the timestamp of the first event could act as a sufficient estimate of 
the resource's creation date. In this paper, we investigate 
using those traces on the web to estimate the creation date of the published resource. Finally, we propose an implementation to this tool based on our analysis to be utilized by researchers.

\section{Related Work}

The problem of estimating the age of web resources has been visited before, but from a different angle. Jatowt et al. investigated the age of web content posted in dynamic pages \cite{Jatowt:2007:DAP:1316902.1316925}. 
They utilized a multiple binary search algorithm to extract the first time the content of a certain DOM component of the page 
started to appear within the page in the archives. They analyzed multiple versions of the web page provided by the public archives. After breaking down the page to multiple DOM components, the archived versions 
were explored using binary search for the first existence of each of these components. The timestamp of this 
first appearance is recorded indicating an estimate for when the enclosed web content, within each component, was created. This approach, relies 
on the archiving coverage of the web provided by the public archives, and the temporal difference between when 
the content's creation date and the time it was crawled and archived. This period of time could range from a few hours in heavily archived pages, up to more than a year in other cases.

To access and analyze the public archives we utilized the Memento framework which facilitated the navigation between the current and the past web \cite{nelson:memento:tr}. We investigated web archival coverage while 
estimating how much of the web is archived \cite{Ainsworth:2011:MWA:1998076.1998100}. In our experiment, we sampled the web forming four different data sets extracted from four 
sources. We found that the amount of the web currently 
archived (or having at least one accessible past version in the public web archives) is highly correlated to 
where the web resource resides. Accordingly, the percentage of coverage ranges from 16\% to 79\%.

This would be the case for long standing resources that exist on the web at the time of archiving. Our recent study, investigating resources related to multiple historical events since 2009, showed that the 
published resources are at continuous risk of disappearance and within the first year of publishing about 11\% 
disappear \cite{longTPDL2012:Losing}. This is important if the resource whose age we wish to estimate existed on the web only briefly. This disappearance event might occur prior to the first archival crawl, resulting 
in complete unattainability of the resource.

An investigation on the web resource itself mining for timestamps in the published content was conducted by Inoue and Tajima \cite{Inoue:2012:NRD:2169095.2169098}. They analyzed web pages for timestamp embedded 
by content management systems (CMS). This approach supports the most popular date formats but could suffer from ambiguity due to dates the mix in the month versus day order in the UK format 
versus in the US one. The authors applied different techniques in attempts to solve this ambiguity. As accurate the results of this approach could be, it still remains specific to CMSs and highly reliant on 
the content itself, reducing its generality.

We propose analyzing different other sources and services to mine for the first appearance of the resource. These services vary in reliability and the results they provide which demanded that we conduct an 
evaluation of each of the services we used and investigate the amount of accuracy lost upon the failure of each service. It is worth noting that McCown and Nelson conducted an experiment to gauge the difference between 
what some services like Google Search might provide from both their API versus the web interface \cite{1255237}. They found a significant difference in the results from both sources. Similarly, Klein conducted a study analyzing the results from using the delicious.com API vs. screen scraping the web interface \cite{kleindelicious}. He proved that screen scraping provided better results than utilizing the 
API, which we considered in our analysis.

\section{Age Estimation Methods}
There are three reasons we cannot use just the web archives to estimate 
the creation date.  First, not all pages are archived.  Second, 
there is often a considerable delay between when the page first 
appeared and when the page was crawled and archived.  Third, web 
archives often quarantine the release of their holdings until after a 
certain amount of time has passed (sometimes 6--12 months). 

These three major deficiencies discourage the use of the web archives solely in estimating an accurate creation date timestamp for web resources. In the following sections, we investigate several other sources 
that explore different areas to uncover the traces of the web resources. Utilizing the best of a range of methods since we cannot rely on one method alone, we build a module that gathers this information and provides a 
collectively estimation of the creation date of the resource. Figure \ref{fig:timeline} illustrates the methodology of the age estimation process with respect to the timeline of the resource. 

\subsection{Resource and Server Analysis}
Prior to investigating any of the web traces we return back to the basics, to the resource itself. We send a request for headers to the hosting web server and parse the output. We search for the existence of last 
modified date response header and parse the timestamp associated if it exists. We use the \texttt{curl} command to request the headers as shown in figure \ref{fig:0}. We also note the timestamp obtained from the headers can 
have errors as demonstrated in a study of the quality of etags and last-modified datestamps by Clausen \cite{Clausen04}.
\begin{figure}[ht]
\textbf{\texttt{curl -I http://ws-dl.blogspot.com/2012/02/2012-02-11-losing-my-\linebreak
revolution-year.html}}
\vskip -5mm
\begin{center}
\begin{verbatim}

HTTP/1.1 200 OK
Content-Type: text/html; charset=UTF-8
Expires: Sat, 02 Mar 2013 04:04:09 GMT
Date: Sat, 02 Mar 2013 04:04:09 GMT
Cache-Control: private, max-age=0
Last-Modified: Wed, 27 Feb 2013 17:27:20 GMT
ETag: "473ba56b-fd4a-4778-b721-3eabdd34154e"
X-Content-Type-Options: nosniff
X-XSS-Protection: 1; mode=block
Content-Length: 0
Server: GSE
\end{verbatim}
\caption{HTTP response headers displaying last modified date field}
\label{fig:0}
\end{center}
\end{figure}
\subsection{Backlinks Analysis}
Typically, we think of backlinks as discoverable through search engines. In the next sections we 
explore the different forms of backlinks and how we can utilize them in our investigation.

\subsubsection{Search Engine Backlinks}
Firstly, a backlink refers to the link created on a page \textit{A} referring to the intended page \textit{B}. Page \textit{A} is considered a backlink of \textit{B}. If Page \textit{A} is static and never 
changed this means that it was created at point in time following the creation of \textit{B}, could be by minutes or years. 
If page \textit{A} was change-prone and had several versions, the first appearance of the link to page \textit{B} on \textit{A} could trigger the same event indicating that that it happened also at a point in time following the 
creation of \textit{B}. If we can search the different versions of A throughout time we can estimate this backlink timestamp.

To accomplish this, we utilized Google API\footnote{https://developers.google.com/custom-search/v1/overview} in extracting the backlinks of the URI. Note that Google API is known to under-report backlinks as shown by McCown and Nelson \cite{1242763}. To explore the multiple versions of each of the backlinks we utilize the Memento framework in accessing the multiple 
public archives available \cite{nelson:memento:tr}. For each backlink we extract its corresponding timemaps. We use binary search to discover in the time maps the first appearance of the link to the investigated resource in the backlink pages. Using binary search ensures the speedy performance of this section of the age estimating 
module. With the backlink having the most archived snapshots (CNN.com > 23,000 mementos), the process took less than 15 iterations accessing the web archives. The minimal of the first appearance 
timestamps from all the backlinks is selected as the estimated backlink creation date. Similarly, this date can act as a good estimation to the creation date of the resource.

\subsubsection{Social Media Backlinks}
Twitter enables users 
to associate a link with their tweeted text, technically creating a backlink to the shared resource. When a user creates a web resource and publicizes it on their social network, by tweeting a link 
to it or posting it on their Facebook account, they create backlinks to their resource. Typically, these backlinks are not accessible via a search engine. The more popular the user and the more the resource gets retweeted or shared, the more backlinks the original 
resource gains increasing its rank and discoverability in search engines.

\begin{figure}[htb]
        \centering
        \subfigure[Resource published at time $t_{creation} = 2012:02:11$. ]{
        \includegraphics[scale=0.5]{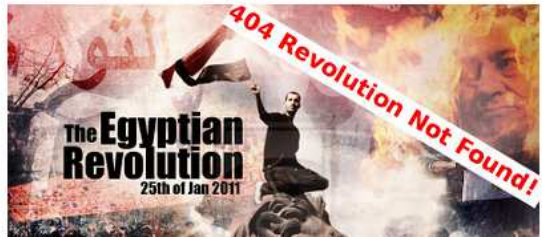}
        \label{fig:1}
        }
        \subfigure[A tweet posted referencing the resource at time $t_{post} = 2012:02:12T06:33:00$. ]{
        \includegraphics[scale=0.5]{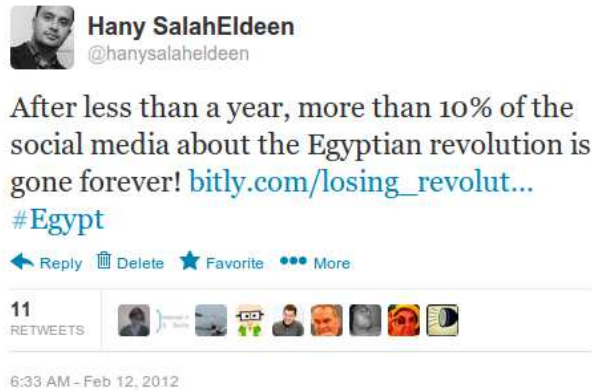}
        \label{fig:2}
        }
        \caption{A published resource and a corresponding social activity.}
        \label{fig:example}
\end{figure}

To elaborate, we examine the following scenario. A resource has been created at time $t_{creation}$, as shown in fig \ref{fig:1} and shortly after a social media post, or a tweet, has been published referring 
to the resource at time $t_{post} = 2012:02:12$ as shown in fig \ref{fig:2}. This new time $t_{post} = 2012:02:12 06:33:00$, could act as a fairly close estimate to the creation date of the post 
with a tolerable margin of error of minutes in some cases between the original $t_{creation}$ and $t_{post}$. 

Given this scenario, tweets inherently are published with a creation/posting date which makes it easier to extract. The task remaining is to find the tweets that were published with the targeted 
resource embedded in the text with incorporating all the shortened versions of the URI as well. Twitter's timeline search facility and its API both provide results of a maximum of 9 days from the 
current day \cite{twittersearch}. Accordingly, we utilize another service, Topsy.com, that enables the user to search for a certain URI and get the latest tweets that incorporated it and the 
influential users sharing it. Topsy's Otter API provides up to 500 of the most recent tweets published embedded a link to the resource and the total number of tweets ever published. Except 
for the highly popular resources, the 500 tweets limit is often sufficient for most resources. The tweets are collected and the corresponding posting timestamps are extracted. The minimum 
of these timestamps acts as an indication of the first time the resource was tweeted. This timestamp in turn signifies the intended $t_{post}$ mentioned earlier.

Another form backlinks could take is URI shortnening. Currently, there are hundreds of services that enables the user to create a short URI that references another longer URI and acts as an alias to it for easier 
dissemination on the web. Shortened URIs could be used for the purposes of customizing the URI or for monitoring the resource by logging the amount of times the short URI have been 
dereferenced or clicked \cite{Antoniades:2011:WWS:1963405.1963505}. Some services, like Bitly, can provide 
the users with a lookup capability for long URIs. When a URI is shortened for the first time by a non logged-in user, it creates an aggregate public short URI that is public to everyone. When 
other unauthenticated users attempt to shorten the same URI it provides the original first aggregated short URI. For every logged-in user, the service provides the possibility to create another personal shortened 
URI. For our purposes we lookup the aggregated short URI indicating the first time the resource's URI have been shortened by this service and from that we query the service once more for the short URI 
creation timestamp. Bitly has been used as the official automatic shortener for years by Twitter before they replaced it with their own shortener. Similarly to the previous backlinks method we mine Bitly 
for those creation timestamps and use them as an estimate of the creation date of the resource, assuming the author shortens and shares the resource's URI shortly after publishing it.

\begin{table*}[ht]
\begin{center}
\resizebox{\textwidth}{!}{
\begin{tabular}{|l|l|l|l|l|}
\cline{2-5}
\multicolumn{1}{c|}{}&\multirow{2}{*}{\textbf{Data Sources}} & \textbf{Resources} & \textbf{Sampled} & \textbf{Timestamp Allocation}  \\
\multicolumn{1}{c|}{}&& \textbf{Collected} & \textbf{Resources} & \textbf{Method}  \\ \hline
\multirow{5}{*}{\rotatebox[origin=c]{90}{\textbf{News Sites}}}&news.Google.com&29,154&100 & XML sitemap\\ \cline{2-5}
&BBC.co.uk&3,703&100 & Page Scraping\\ \cline{2-5}
&CNN.com&18,519&100 & Page Scraping\\ \cline{2-5}
&news.Yahoo.com&34,588&100 & XML sitemap\\ \cline{2-5}
&theHollywoodGossip.com&6,859&100 & Page Scraping\\ \hline\hline
\multirow{5}{*}{\rotatebox[origin=c]{90}{\textbf{Social Sites}}}&Pinterest.com&55,463&100 & RSS feed\\ \cline{2-5}
&Tumblr.com&52,513&100 & RSS feed\\ \cline{2-5}
&Youtube.com&78,000&100 &Search API \\ \cline{2-5}
&WordPress.com&2,405,901&100 & Atom feed\\ \cline{2-5}
&Blogger.com&32,417&100  & Atom feed\\ \hline \hline
\multicolumn{1}{c|}{}&Alexa.com Top Domains&167&100 & Page Scraping \& Who.is service\\ \cline{2-5}
\multicolumn{1}{c|}{}&Manual Extraction&100&100 & Manual inspection\\  \cline{2-5}
\multicolumn{2}{r|}{\textbf{Total:}}&2,717,384&\textbf{1,200} & \multicolumn{1}{c}{}\\  \cline{3-4}
\end{tabular}
}
\vskip 3mm
\caption{\label{tab:1}The resources extracted with timestamps from the web forming the gold standard dataset.}
\end{center}
\end{table*}
\subsection{Archiving Analysis}
The most straightforward approach used in the age estimation module is the web archives analysis. We utilize the Memento framework to obtain the timemap of the resource and from which we obtain the memento 
datetime for each and then extract the least one indicating the first memento captured. Note that memento datetime is the time of capture at the web archive and is not equivalent to last modified or creation dates \cite{nelson2011memento}. 
In some cases, the original headers in some mementos include the original last modified dates, but all of them have the memento date time fields. We extract 
each of those fields, parse the corresponding dates, and pick the lowest of which. An extra date range filter was added to avoid dates prior to 1995, before the Internet Archive began archiving, or more than the current timestamp.

\subsection{Search Engine Indexing Analysis}
The final approach is to investigate the search engines and extract the last crawled date. Except for the highly active and dynamic web pages, the resources get crawled once and get marked as such to 
prevent unnecessary re-crawling. News sites article pages, blogs, and videos are the most encountered examples of this. The idea is to use the search engines' APIs to extract this last crawled date and 
utilize it as an estimate of the creation date. This approach is effective due to the relatively short period of time between publishing a resource and its discovery by search engine crawlers. We use Google's 
search API and modify it to show the results from the last 15 years accompanied by the first crawl date. Unfortunately this approach does not give time granularity (HH:MM:SS), just dates (YYYY:MM:DD).

\section{Estimated Age Verification}
To validate an implementation of the methods described above, we collect a gold standard dataset from different sources which we can extract the real publishing timestamps. This could be 
done by parsing feeds, parsing web templates, and other methods. In the next sections we illustrate each of the sources utilized and explain the extraction process.

\subsection{Gold Standard Data Collection}
Two important factors were crucial in the data collection process: The quality of the timestamps extracted, and the variety of the sources to reduce any bias in the experiment. 
Thus, we divide data into four categories. Table \ref{tab:1} summarizes the four categories.

\subsubsection{News Sites}
Each article is associated with a timestamp in a known template that can be parsed and extracted. The articles are also usually easily accessible through RSS and Atom feeds or XML sitemaps. For each of the news sites under investigation we extracted as 
many resources as possible then randomly downsized the sample.

\subsubsection{Social Media and Blogs}
To increase the variety of the gold standard dataset we investigate five different social media sources. These selected sources are highly popular, and it is possible to extract accurate publishing timestamps. 
As those sources are tightly coupled with the degree of popularity and to avoid the bias resulting from this popularity we randomly extract as many resources as 
possible from the indexes, feeds, and sitemaps and do not rely solely on the most famous blogs or most shared tumblr posts. Furthermore, we randomly and uniformly sample 
each collection to reduce its size for our experiment.

\begin{table*}[ht]
\begin{center}
\resizebox{\textwidth}{!}{
\begin{tabular}{|c||c|c|c|c||c|c|}
\hline
Age &  Resources Found&  Percentage& \multicolumn{2}{c||}{Contribution}  &  \multicolumn{2}{c|}{Area Under Curve} \\ \cline{4-7}
Estimation & By The Best & Of Resources &  Resources &  Percentage &  AUC & Percentage lost\\ 
Method & Estimate Method & Found & Contributed &  Contributed &  & in AUC\\ \hline\hline
\textbf{Bitly} & 96 & 10.55\% & 554 & 46.21\% & 758.73 & 0.51\%\\ \hline
\textbf{Google} & 370 & 40.66\% & 709 & 59.13\% & 742.52 & 2.64\%\\ \hline
\textbf{Topsy} & 236 & 25.93\% & 632 & 52.71\% & 720.61 & 5.51\%\\ \hline
\textbf{Archives} & 152 & 16.70\% & 578 & 48.21\% & 741.23 & 2.81\%\\ \hline
\textbf{Backlinks} & 3 & 0.33\% & 180 & 15.01\% & 762.64 & 0\%\\ \hline
\textbf{Last Modified} & 53 & 5.82\% & 134 & 11.18\% & 725.59 & 4.86\%\\ \hline\hline
\textbf{Total Estimate} & \textbf{910} & \textbf{75.90\%} & 1199 & 100\% & 762.64 & 0\%\\ \hline
\end{tabular}
}
\caption{\label{tab:auc}Results of testing the gold standard dataset against the six age estimation methods (n=1200).}
\end{center}
\end{table*}

\subsubsection{Long Standing Domains}
So as not to limit our gold standard dataset to low level articles, blogs, or posts only, we incorporated top level, long-standing domains. To extract a list of those domains we mined Alexa.com for 
the list of the top 500 sites\footnote{http://www.alexa.com/topsites}. 
This list of sites was in turn investigated for the DNS registry dates using one of the DNS lookup tools available online. A 
final set of 100 was randomly selected from the resolved sites and added to the gold standard dataset.\

\subsubsection{Manual Random Extraction}
Finally, we randomly select a set of 100 URIs that we can visually identify the timestamp somewhere on the page itself. 
These URIs were selected empirically using random walks on the web. The 10 URIs analyzed in \cite{Jatowt:2007:DAP:1316902.1316925} included within these 100 URIs as well. 
The corresponding true value of the creation timestamp for each of the 10 URIs is the one provided in their analysis.

\subsection{Experimental Analysis}
The collected dataset of 1,200 data points is tested against the developed implementation of the carbon dating methods and the results are recorded. Since the data points are collected from different sources, 
the granularity varies in some cases, as well as the corresponding time zones. To be consistent, each real creation date timestamp $t_{real}$ is transformed from the corresponding extracted timestamp to Coordinated Universal Time (UTC) and the 
granularity for all the timestamped have been set to be a day. Each data point has a real creation date in the ISO 8601 date format without the time portion (e.g., YYYY:MM:DD). Similarly, the extracted estimations 
were processed in the same manner and recorded.

For each method, we record the estimated timestamp $t_{method}$ and the temporal delta $\Delta t_{method}$ between the 
estimated timestamp $t_{method}$ and the actual one $t_{real}$ as shown in equation \ref{eq:1}. Collectively, we calculate the best estimated timestamp $t_{estimated}$ as in equation \ref{eq:2}, the closest delta between all the methods $\Delta t_{least}$ and 
the real timestamp $t_{real}$ as shown in equation \ref{eq:3}, and the method that provided this best estimate.

\begin{equation}
\Delta t_{method} = |t_{real}-t_{method}|
 \label{eq:1}
\end{equation}
\begin{equation}
t_{estimated} = min(t_{method})
 \label{eq:2}
\end{equation}
\begin{equation}
\Delta t_{least} = |t_{real}-t_{estimated}|
 \label{eq:3}
\end{equation}

\begin{figure*}
\centering
\includegraphics[width=\textwidth]{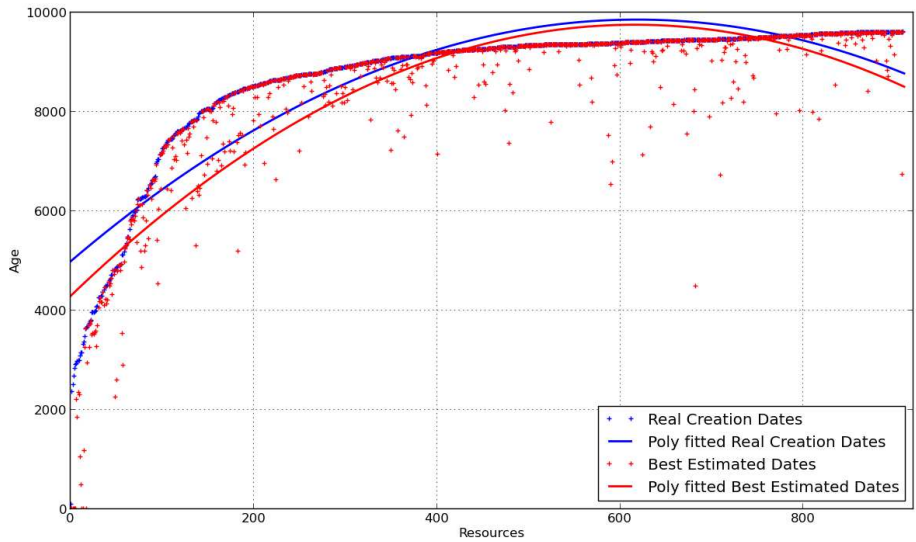}
\caption{The polynomial fitted curve corresponding to the real creation dates against the estimated creation dates from the module AUC = 762.64.}
\label{fig:fitted}
\end{figure*}

Table \ref{tab:auc} shows the outcomes of the experiment. The numbers indicate how many times a resource provided the closest timestamp to the real one. 
It also shows that for 290 resources, the module failed to provide a single creation date estimate (24.90\%).
\section{Evaluation}

As our age estimation module relies on other services to function (e.g., Bitly, Topsy, Google, Web Archives); the next step is to measure the effect 
of each of the six different age estimation methods and to gauge the consequences resulting in failure to obtain results from each. For each resource we get the resulting best estimation and calculate the distance between 
it and the real creation date. We set the granularity of the delta to be in days to match the real dates in the gold standard dataset. To elaborate, if the resource was created on a certain date and the estimation module 
returned a timestamp on the same day we declare a match and in this case $\Delta t_{least}$ = 0. To measure the accuracy of estimation, 393 resources out of 1200 (32.78\%) returned $\Delta t_{least}$ = 0 indicating a perfect estimation. For all 
the resources, we sort the resulting deltas and plot them. We calculate the area under the curve using the 
composite trapezoidal rule and the composite Simpon's rule with X-axis spacing of 0.0001 units. We take the average of both approximations to represent the area under the curve (AUC). Semantically, this area signifies the 
error resulting from the estimation process. Ideally, if the module produced a perfect match to the real dates, AUC = 0. Table \ref{tab:auc} shows that the AUC using the best lowest estimate of 
all the six methods is 762.64. Disabling each method one by one and measuring the AUC indicates the resultant error corresponding to the absence of the disabled method accordingly. The table shows that using or disabling 
the use of backlinks barely affected the results. Disabling the Bitly services or the Google search index query affected the results slightly (0.51\% and 2.64\% respectively). While disabling any of the public archives query, or the social backlinks in 
Topsy and the extraction of the last modified date if exists hugely affects the results increasing the error tremendously.

\begin{figure*}[ht!]
     \begin{center}
      \centering
        \subfigure[Without Bitly, AUC=758.73]{%
            \label{fig:first}
            \includegraphics[width=0.5\textwidth]{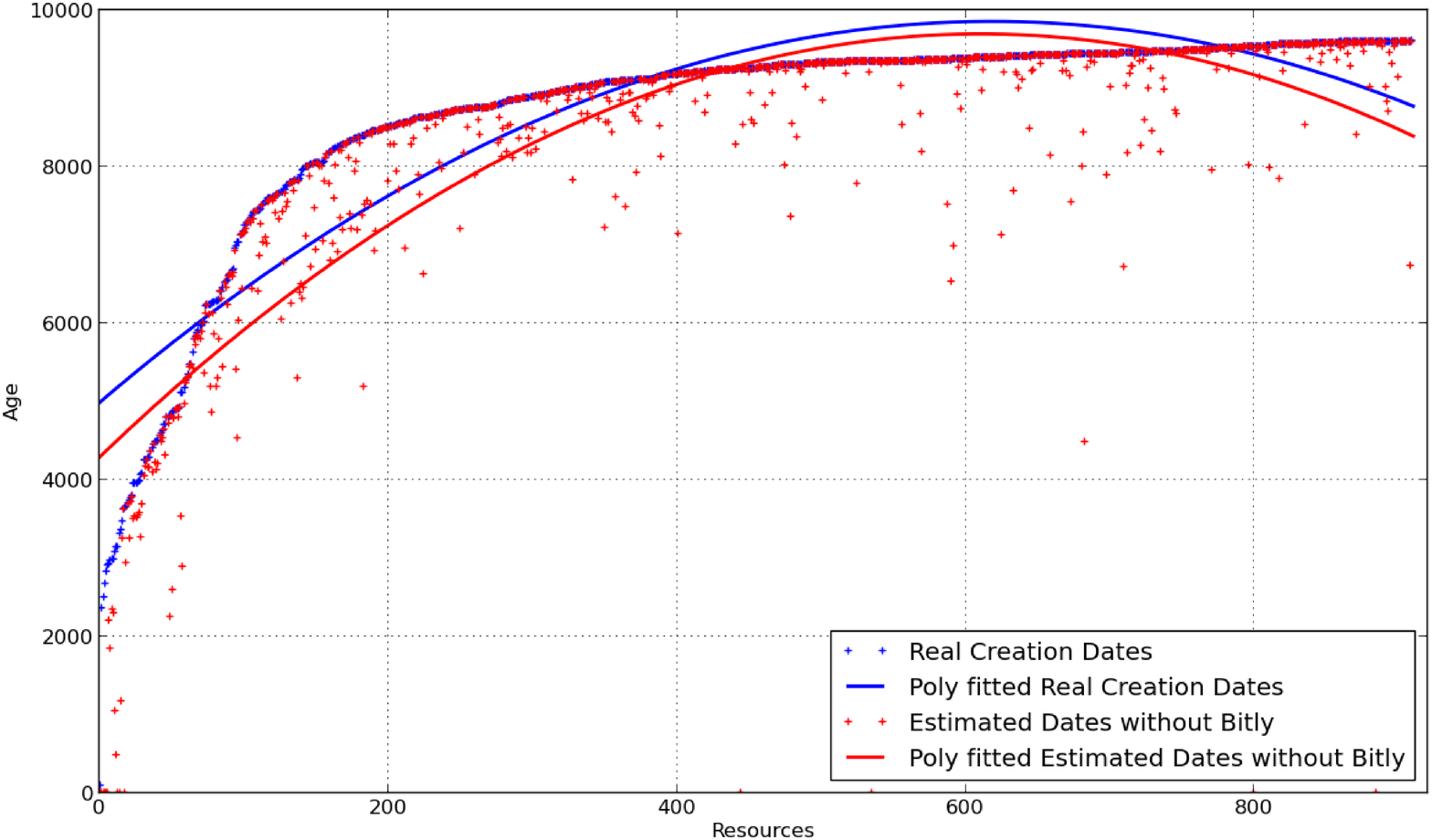}
        }%
        \subfigure[Without Google, AUC=742.52]{%
           \label{fig:second}
           \includegraphics[width=0.5\textwidth]{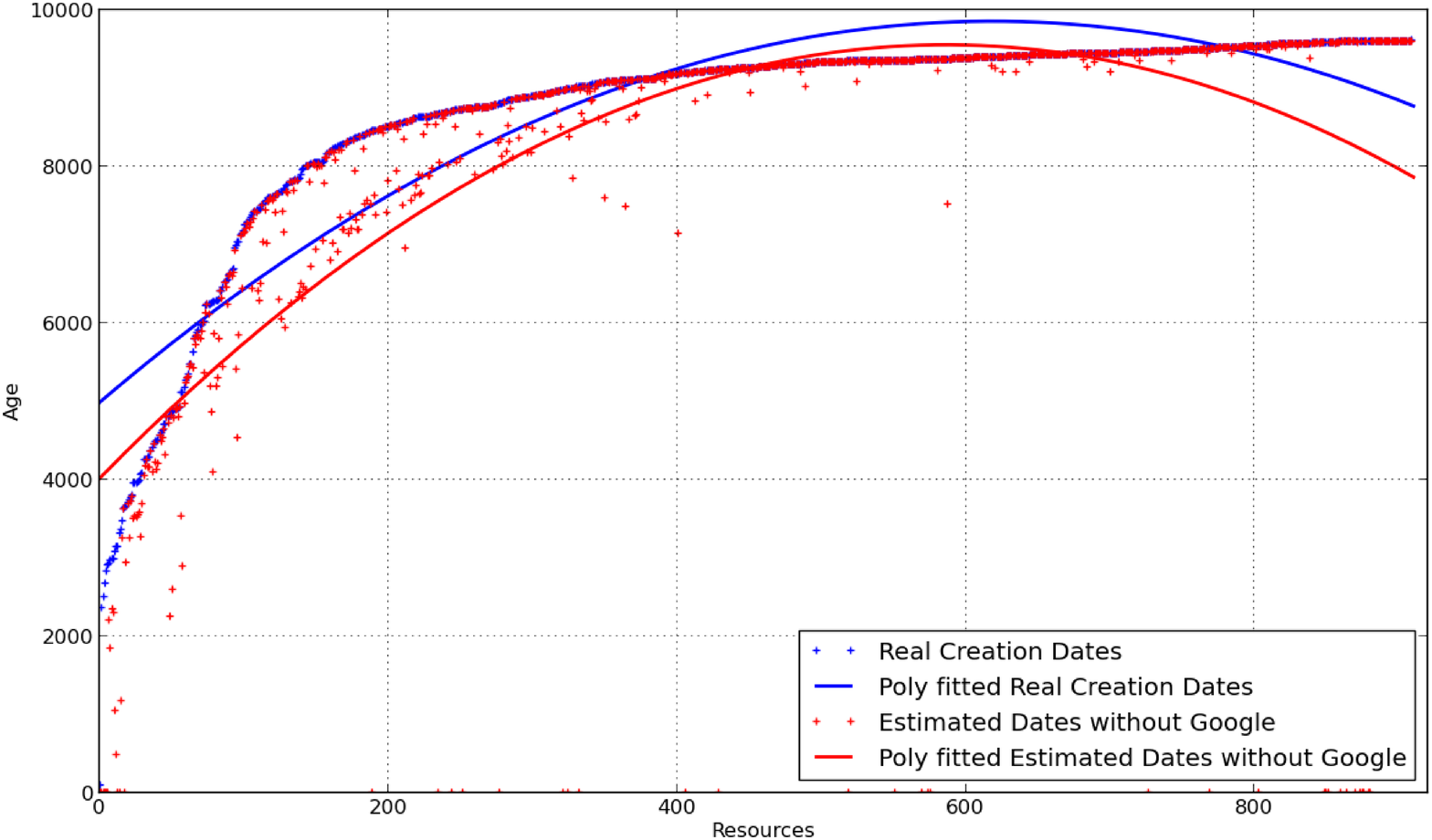}
        }\\ 
        \subfigure[Without Topsy, AUC=720.61]{%
            \label{fig:third}
            \includegraphics[width=0.5\textwidth]{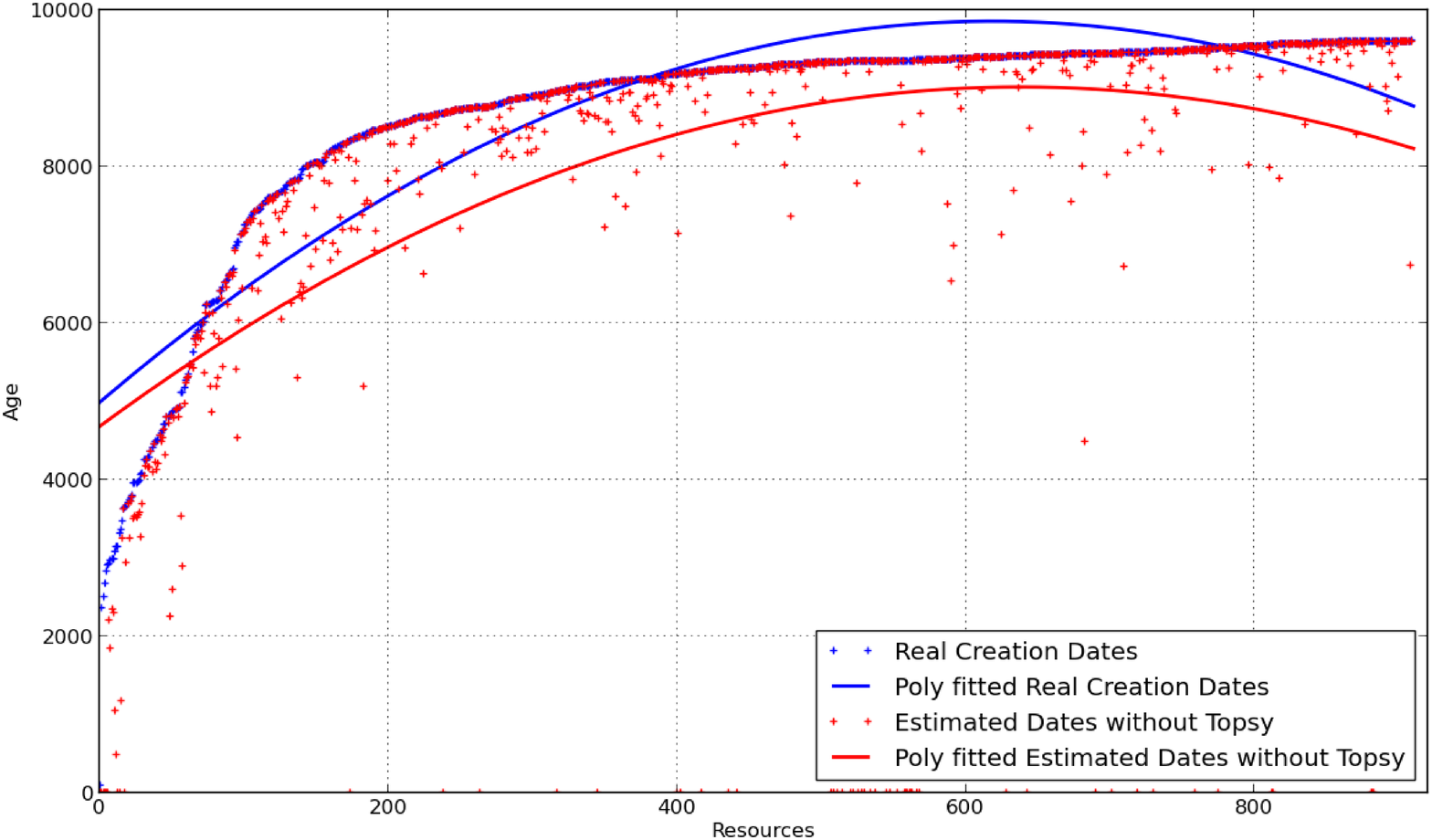}
        }%
        \subfigure[Without Last Modified, AUC=725.59]{%
            \label{fig:fourth}
            \includegraphics[width=0.5\textwidth]{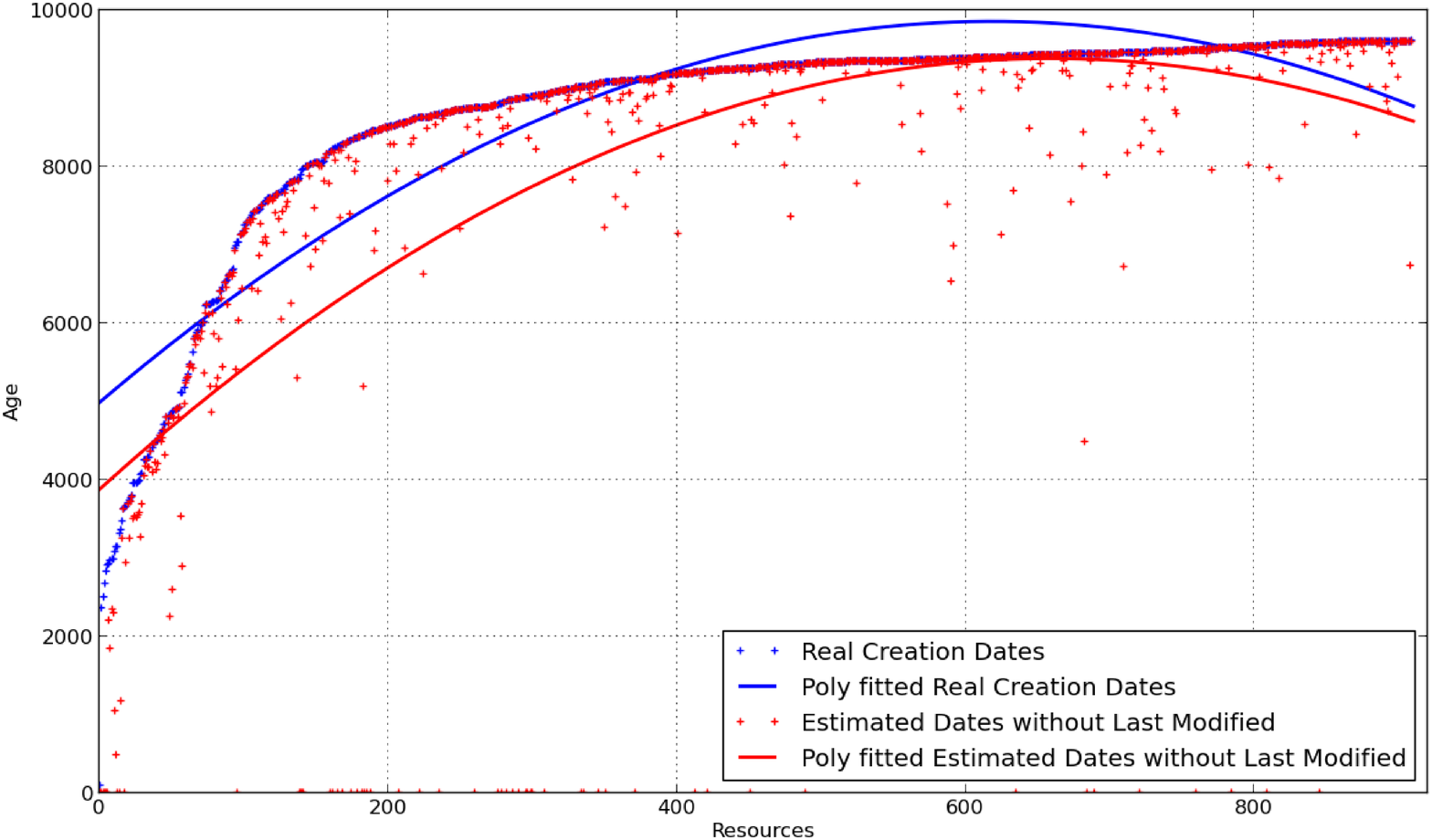}
        }\\ 
        \subfigure[Without Archives, AUC=741.23]{%
            \label{fig:fifth}
            \includegraphics[width=0.5\textwidth]{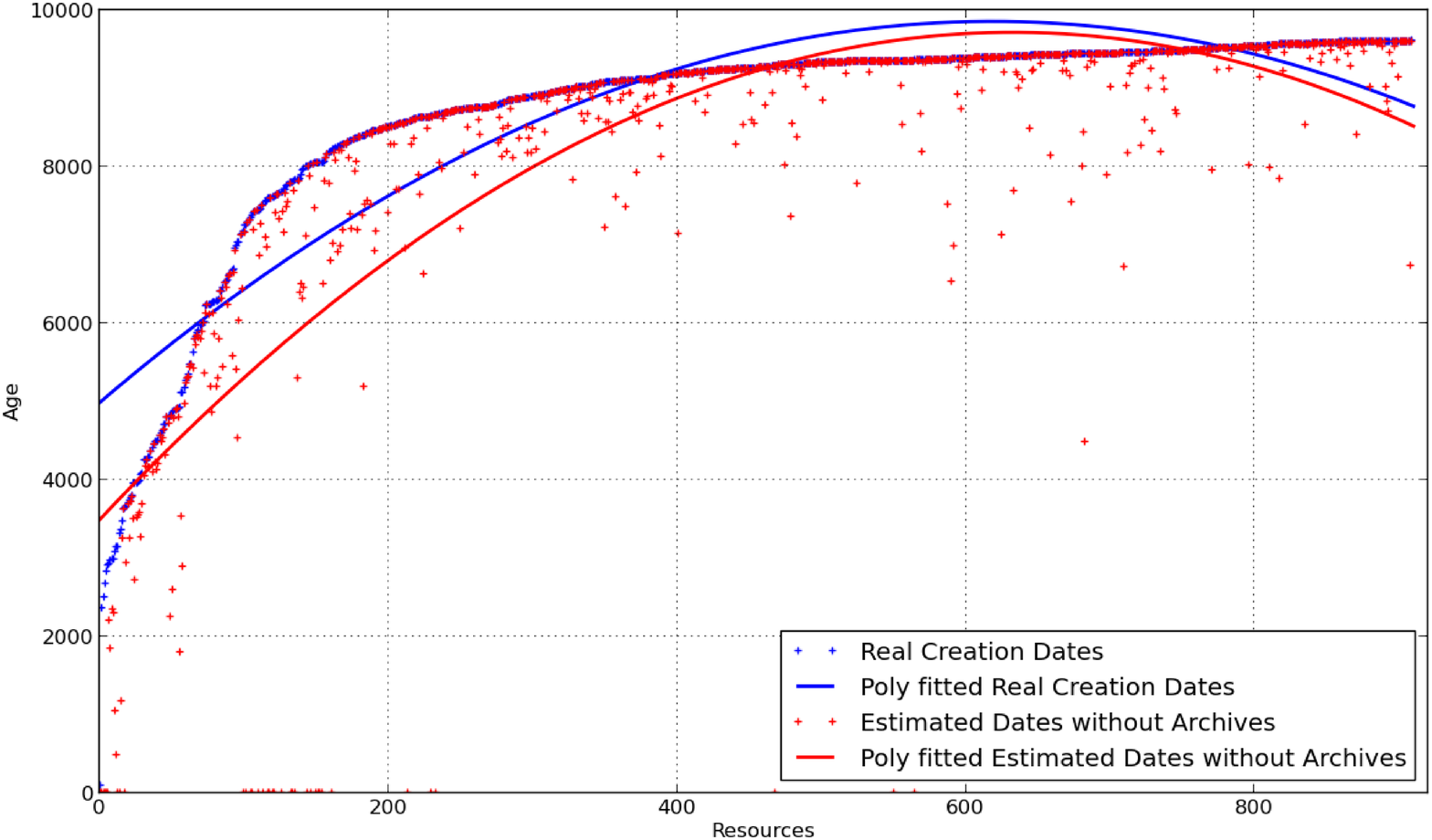}
        }%
    \end{center}
    \caption{The polynomial fitted curves corresponding to the absence of each method separately.}%
   \label{allfitted}
\end{figure*}

We utilized polynomial fitting functions to fit the values corresponding to the age estimations corresponding to each URI. Figure \ref{fig:fitted} shows the polynomial curve of the second degree used in fitting the 
real creation times stamps of the gold standard dataset. Figure \ref{allfitted} shows the fitted curve resulting from removing each of the methods one by one. Each of the curves signifies an estimate of the best 
the other methods could provide. The further the estimated curve is from the real one the less accurate this estimation would be.

\begin{figure}[ht]
\textbf{\texttt{curl -i http://cd.cs.odu.edu/cd/http://www.mementoweb.org}}
\vskip -5mm
\begin{verbatim}

HTTP/1.0 200 OK
Date: Fri, 01 Mar 2013 04:44:47 GMT
Server: WSGIServer/0.1 Python/2.6.5
Content-Length: 550
Content-Type: application/json; charset=UTF-8

{
  "URI": "http://www.mementoweb.org",
  "Estimated Creation Date": "2009-09-30T11:58:25",
  "Last Modified": "2012-04-20T21:52:07",
  "Bitly": "2011-03-24T10:44:12",
  "Topsy.com": "2009-11-09T20:53:20",
  "Backlinks": "2011-01-16T21:42:12",
  "Google.com": "2009-11-16",
  "Archives": {
    "Earliest": "2009-09-30T11:58:25",
    "By Archive": {
      "wayback.archive-it.org": "2009-09-30T11:58:25",
      "api.wayback.archive.org": "2009-09-30T11:58:25",
      "webarchive.nationalarchives.gov.uk": "2010-04-02T00:00:00"
    }
  }
}
\end{verbatim}
\caption{JSON Object resulting from the Carbon Date API}
\label{fig:json}
\end{figure}

\section{Application: Carbon Date API}
After validating the accuracy of the developed module the next step was to openly provide age estimation as a web service. To fulfill this goal, we created ``\textbf{\textit{Carbon Date}}'', a web based 
age estimation API. To use the API, simply concatenate the URI of the desired resource to the following path:\\
\textit{http://cd.cs.odu.edu/cd/} and the resulting JSON object would be similar to the one illustrated in figure \ref{fig:json}.

\section{Conclusions}
Estimating the age of web resources is essential for many areas of research. 
Previous research investigated the use of public archives as a point of reference to when the content of a certain page appeared. In this study, we investigated several other possibilities in 
estimating the accurate age of a resource including social backlinks (social posts and shortened URIs), search engine backlinks, search engine last crawl date, the resource last modifed date, the first 
appearance of the link to the resource in its backlinks sites, and the archival first crawl timestamp. We also incorporated the minimum of the original headers last modified date, 
and the Memento-Datetime HTTP response header. All of these methods combined, where we select the oldest resulting timestamp, proved to provide an accurate estimation to the creation date upon evaluating it against a gold 
standard dataset of 1200 web pages of known publishing/posting dates. We succeeded in obtaining an estimated creation date to 910 resources out of the 1200 in the dataset (75.90\%). 40\% of the closest estimated dates were obtained from Google, 
Topsy came in second with 26\%, followed by the public archives, Bitly, and Last Modified header with 17\%, 11\%, and 6\% respectively. Using the backlinks yielded only 3 closest creation dates proving its insignificance. We also simulate the failure of each of the six services one at a time and calculated the resulting loss in accuracy. We show that the 
social media existence (Topsy), the archival existence (Archives) and the last modified date if it exists, are the strongest contributers to the age estimation module respectively.

\section{Acknowledgments}
This work was supported in part by the Library of Congress and NSF IIS-1009392.

\end{document}